\documentclass{article}
\usepackage{emulateapj,timesfonts}

\def\***#1{{\sc #1}}
\def\ROSAT{{\em ROSAT}}
\def\r180{r_{180}}

\renewcommand{\arraystretch}{1.2}

\begin{document}

\lefthead{VIKHLININ ET AL.}
\righthead{OUTER REGIONS OF CLUSTERS}
\submitted{Submitted to the Astrophysical Journal, Novermber 30, 1998;
  revised May 17, 1999}

\title{OUTER REGIONS OF THE CLUSTER GASEOUS ATMOSPHERES}
\author{A. Vikhlinin, W. Forman, C. Jones}
\affil{Harvard-Smithsonian Center for Astrophysics, 60 Garden St.,
  Cambridge, MA 02138;\\
  avikhlinin wforman cjones @cfa.harvard.edu}

\begin{abstract}
  
  We present a systematic study of the hot gas distribution in the outer
  regions of regular clusters using \emph{ROSAT} PSPC data. Outside the
  cooling flow region, the $\beta$-model describes the observed surface
  brightness closely, but not precisely. Between 0.3 and 1 virial radii, the
  profiles are characterized by a power law with slope, expressed in terms
  of the $\beta$ parameter, in the range $\beta=0.65$ to $0.85$. The values
  of $\beta$ in this range of radii are typically larger by $\approx0.05$
  than those derived from the global fit. There is a mild trend for the
  slope to increase with temperature, from $\langle\beta\rangle\approx0.68$
  for 3~keV clusters to $\approx 0.8$ for 10~keV clusters; however, even at
  high temperatures there are clusters with flat gas profiles, $\beta<0.7$.
  Our values of $\beta$ at large radius are systematically higher, and the
  trend of $\beta$ with temperature is weaker than was previously found; the
  most likely explanation is that earlier studies were affected by an
  incomplete exclusion of the central cooling flow regions. For our regular
  clusters, the gas distribution at large radii is quite close to
  spherically symmetric and this is shown not to be an artifact of the
  sample selection.  The gas density profiles are very similar when compared
  in the units of cluster virial radius.  The radius of fixed mean gas
  overdensity 1000 (corresponding to the dark matter overdensity 200 for
  $\Omega=0.2$) shows a tight correlation with temperature, $R\sim T^{0.5}$,
  as expected from the virial theorem for clusters with the universal gas
  fraction.  At a given temperature, the \emph{rms} scatter of the gas
  overdensity radius is only $\approx 7\%$ which translates into a 20\%
  scatter of the gas mass fraction, including statistical scatter due to
  measurement uncertainties.
  
\end{abstract}


\begin{table*}
  \footnotesize
  \begin{center}
    \caption{Cluster Sample}
\begin{tabular}{p{1.75cm}lrcccrc}
\hline
\hline
\multicolumn{1}{c}{Name} & 
\multicolumn{1}{c}{$z$}  &
\multicolumn{1}{c}{$T$} &
\multicolumn{1}{c}{Ref.} &
\multicolumn{1}{c}{$r_{\rm 180}(T)$} &
\multicolumn{1}{c}{$R_{\rm cool}$} &
\multicolumn{1}{c}{$N_H$} &
\multicolumn{1}{c}{Note} \\
 & 
 &
\multicolumn{1}{c}{(keV)} &
 &
(arcmin) &
(arcmin) &
\multicolumn{1}{c}{($10^{20}\,$cm$^{-2}$)} &
 \\
\hline
\\ \multicolumn{7}{c}{\sc Cooling flow clusters}\\
   2A0335\dotfill & $0.035$ & $3.0\pm0.1$ & 1 & 37.3 & 3.8$^{a}$ & 18.12\phantom{DD} &            \\
      A85\dotfill & $0.052$ & $6.9\pm0.2$ & 2 & 39.0 & 1.8$^{a}$ & 3.36\phantom{DD} & ? \\
     A133\dotfill & $0.060$ & $3.8\pm0.8$ & 3 & 25.3 & 1.6$^{b}$ & 1.58\phantom{DD} &            \\
     A478\dotfill & $0.088$ & $8.4\pm0.7$ & 2 & 27.0 & 1.5$^{a}$ & 15.08\phantom{DD} &            \\
     A496\dotfill & $0.033$ & $4.9\pm0.1$ & 4 & 50.5 & 2.0$^{a}$ & 4.79\phantom{DD} &            \\
     A644\dotfill & $0.071$ & $8.1\pm0.5$ & 2 & 32.0 & 1.3$^{a}$ & 6.52\phantom{DD} &            \\
     A780\dotfill & $0.057$ & $4.3\pm0.2$ & 2 & 28.4 & 1.8$^{a}$ & 4.84\phantom{DD} &            \\
    A1651\dotfill & $0.085$ & $6.3\pm0.3$ & 2 & 24.2 & 1.0$^{a}$ & 1.88\phantom{DD} &            \\
    A1689\dotfill & $0.184$ & $9.0\pm0.2$ & 5 & 15.5 & 0.8$^{a}$ & 1.81\phantom{DD} &            \\
    A1795\dotfill & $0.062$ & $7.8\pm0.6$ & 2 & 35.3 & 1.8$^{a}$ & 1.17\phantom{DD} &            \\
    A2029\dotfill & $0.077$ & $9.1\pm0.6$ & 2 & 31.7 & 1.6$^{a}$ & 3.15\phantom{DD} &            \\
    A2052\dotfill & $0.035$ & $3.1\pm0.2$ & 3 & 37.5 & 2.5$^{a}$ & 2.84\phantom{DD} &            \\
    A2063\dotfill & $0.035$ & $4.1\pm0.6$ & 3 & 42.9 & 1.6$^{a}$ & 3.04\phantom{DD} &            \\
    A2142\dotfill & $0.089$ & $9.7\pm0.8$ & 2 & 28.6 & 1.1$^{a}$ & 4.16\phantom{DD} & ? \\
    A2199\dotfill & $0.030$ & $4.8\pm0.1$ & 4 & 54.3 & 2.9$^{a}$ & 0.89\phantom{DD} &            \\
    A2597\dotfill & $0.085$ & $4.4\pm0.3$ & 2 & 20.1 & 1.2$^{a}$ & 2.49\phantom{DD} &            \\
    A2657\dotfill & $0.040$ & $3.7\pm0.2$ & 2 & 36.4 & 1.6$^{b}$ & 5.57\phantom{DD} &            \\
    A2717\dotfill & $0.050$ & $2.2\pm0.5$ & 6 & 22.9 & 1.9$^{c}$ & 1.11\phantom{DD} &            \\
    A3112\dotfill & $0.076$ & $5.3\pm0.5$ & 2 & 26.2 & 1.8$^{a}$ & 2.53\phantom{DD} &            \\
    A3571\dotfill & $0.040$ & $6.9\pm0.1$ & 2 & 49.7 & 1.6$^{a}$ & 4.11\phantom{DD} &            \\
    A4038\dotfill & $0.028$ & $3.3\pm0.8$ & 3 & 47.7 & 2.8$^{a}$ & 1.54\phantom{DD} &            \\
    A4059\dotfill & $0.048$ & $4.4\pm0.2$ & 2 & 33.5 & 2.0$^{a}$ & 1.10\phantom{DD} &            \\
     AWM4\dotfill & $0.032$ & $2.4\pm0.1$ & 1 & 36.0 & 1.0$^{b}$ & 4.99\phantom{DD} &            \\
    MKW3S\dotfill & $0.045$ & $3.7\pm0.1$ & 2 & 32.6 & 2.4$^{a}$ & 3.05\phantom{DD} &            \\
     MKW4\dotfill & $0.020$ & $1.7\pm0.1$ & 1 & 47.9 & 2.1$^{b}$ & 1.88\phantom{DD} & ? \\
\\ \multicolumn{7}{c}{\sc Non-cooling flow clusters}\\
      A21\dotfill & $0.095$ & $5.3\pm1.0$ & 7 & 20.1 & ... & 4.44\phantom{DD} &            \\
     A400\dotfill & $0.024$ & $2.3\pm0.1$ & 1 & 46.8 & ... & 9.39\phantom{DD} & ? \\
     A401\dotfill & $0.074$ & $8.0\pm0.2$ & 2 & 30.6 & 0.7$^{a}$ & 10.16\phantom{DD} &            \\
     A539\dotfill & $0.029$ & $3.2\pm0.1$ & 1 & 46.4 & 0.7$^{b}$ & 12.77\phantom{DD} &            \\
    A1413\dotfill & $0.143$ & $6.7\pm0.2$ & 5 & 16.2 & ... & 1.92\phantom{DD} &            \\
    A2163\dotfill & $0.203$ & $13.9\pm0.6$ & 3 & 18.0 & ... & 12.01\phantom{DD} & ? \\
    A2218\dotfill & $0.171$ & $7.5\pm0.3$ & 5 & 15.0 & 0.4$^{b}$ & 3.14\phantom{DD} &            \\
    A2255\dotfill & $0.080$ & $7.3\pm1.0$ & 3 & 27.3 & ... & 2.53\phantom{DD} &            \\
    A2256\dotfill & $0.058$ & $7.3\pm0.3$ & 2 & 36.4 & ... & 4.07\phantom{DD} & ? \\
    A2382\dotfill & $0.065$ & $2.9\pm0.7$ & 7 & 20.7 & ... & 4.16\phantom{DD} &            \\
    A2462\dotfill & $0.075$ & $2.5\pm0.6$ & 6 & 16.9 & ... & 3.07\phantom{DD} &            \\
    A3301\dotfill & $0.054$ & $3.0\pm0.7$ & 6 & 24.9 & ... & 2.34\phantom{DD} &            \\
    A3391\dotfill & $0.054$ & $5.4\pm0.4$ & 2 & 33.4 & ... & 5.48\phantom{DD} &            \\
  Tri Aus\dotfill & $0.051$ & $9.6\pm0.4$ & 2 & 46.9 & 0.9$^{a}$ & 13.28\phantom{DD} &            \\
\hline
\end{tabular}

    \label{tab:sample}
  \end{center}
$^a$ --- Peres et al.\ (1998), \quad $^b$ --- White et al.\ 1997, 
\quad $^c$ --- our own estimate.

Temperature references: 1 --- Fukuzawa et al.\ (1998), 2 --- Markevitch et
al.\ 1998, 3 --- David et al.\ (1993),  4 --- Markevtch et al.\ (1999), 5
--- Mushotzky \& Scharf (1997), 6 --- our estimate from the $L-T$ correlation,
7 --- Ebeling et al.\ (1996).

Question mark in the last column denotes those clusters with some
substructure in the \emph{ROSAT} image.
\end{table*}

\section{Introduction}

Clusters of galaxies are very important tools for observational cosmology.
Massive clusters form through collapse of a large volume and therefore
thought to contain a fair sample of the Universe in terms of dark matter,
diffuse baryons, and possibly stellar mass. Through the study of the
relative contribution of these components in clusters, one can determine the
average matter density in the Universe as a whole (White et al.\ 1993,
Carlberg et al.\ 1996). 

Most of mass in clusters is in the form of dark matter, observable directly
only through the gravitational distortion of background galaxy images. For
various reasons (sparseness, limited area coverage) lensing observations
still cannot be used for a detailed study of the dark matter distribution in
clusters. Much progress in understanding the dark matter halos of clusters
has been done theoretically, through cosmological numerical simulations.
Properties of simulated clusters in many respects agree with analytic or
semi-analytic theoretical predictions. The mass function of clusters is in
good agreement with that predicted by Press \& Schechter (1974) theory
(Efstathiou et al.\ 1985, Lacey \& Cole 1994). A virialized region is well
defined by $\r180$, a radius within which the mean density is approximately
180 of the critical density (Cole \& Lacey 1996). Simulations predict that
the dark matter density profiles are very similar when the radii are scaled
to $\r180$, the hot gas follows the dark matter distribution at large radii,
and these two components have equal temperature (Navarro, Frenk, \& White
1995). As expected from the virial theorem, the gas temperature in
simulations scales as $M_{180}\propto T^{3/2}$, where $M_{180}$ is the mass
within $\r180$ (Evrard, Metzler, \& Navarro 1996).

Most baryons, i.e.\ observable matter, in clusters are in the form of hot,
X-ray emitting gas. Therefore, most of our direct knowledge about the
structure of clusters comes from X-ray observations. Important cosmological
conclusions derived from X-ray observations of clusters usually rely on
simple theoretical assumptions. For example, a measurement of $\Omega$ from
the baryon fraction in clusters (White et al.\ 1993, David, Jones, \&
Forman 1995, Evrard 1997) requires that cluster baryons are not segregated
with respect to the dark matter.  Measurement of the cosmological parameters
from the evolution of the cluster temperature function (Henry 1997) relies
on the converting of temperature to mass as $M\propto T^{3/2}$. However,
unlike dark matter in simulated clusters, properties of the hot gas inferred
from X-ray observations often deviate from simple theoretical expectations.
For example, if gas were to follow the dark matter of Navarro et al.\ 
(1995), one would observe $\rho_{\rm g} \sim r^{-2.7}$ in the outer cluster
parts (at $r\sim 1\,$Mpc), whereas the gas density profiles inferred from
the {\em Einstein} observatory images are significantly flatter, $\rho_{\rm
  g} \sim r^{-1.8}$ (Jones \& Forman 1984, 1998). The universal density
profile, virial theorem, and the non-segregation of baryons predict the
relation between X-ray luminosity and gas temperature $L\propto T^2$.  The
observed relation is significantly steeper, $L\sim T^{2.6-3}$ (David et al.\ 
1993, Markevitch 1998, Allen \& Fabian 1998). Using the gas temperature
profile measured by {\em ASCA} and assuming that gas is in hydrostatic
equilibrium, Markevitch \& Vikhlinin (1997) derived the mass of A2256 which
was found to be 40\% lower than expected from Evrard et al.\ (1996) scaling.
In fact, the only easily understandable scaling involving cluster baryons
established so far is that between the temperature and galaxy velocity
dispersion $T\propto\sigma^2$ (e.g., Edge \& Stewart 1991).

We demonstrate in this work that the hot gas in clusters does show a scaling
expected from simple theoretical arguments. It is expected that the cluster
virial radius can be defined as a radius of mean overdensity $\approx 180$.
If baryons are not segregated on the global cluster scales, this radius can
be found as a radius of some \emph{baryon} overdensity, i.e.\ determined
observationally. The virial theorem implies that the scaling of this radius
with temperature should be of the form $R\propto T^{1/2}$. Furthermore, if
cluster density profiles are similar, such scaling should be observed for a
range of limiting baryon overdensities. We indeed observe such a relation;
its tightness is comparable to the tightness of similar correlations in
simulated clusters.

We use the values of cosmological parameters $H_0=50$~km~s$^{-1}$~Mpc$^{-1}$
and $q_0=0.5$. The radius of mean gas overdensity $\Delta_g=Y$ relative to
the cosmic baryon density predicted by primordial nucleosynthesis is
referred to as $R_Y$.

\section{Cluster sample}

Our goals require a sample of clusters that are symmetric and that have
high-quality imaging data to large radii.  The present sample includes those
clusters in which the X-ray surface brightness distribution has been mapped
by the \ROSAT\/ PSPC to large radius, i.e.\ those in which the virial
radius, $\r180(T)$, lies within the \ROSAT\/ PSPC field of view. For the
purposes of this work, the virial radius is estimated from the temperature
as $r_{180} = 1.95\, h^{-1}\,\mathrm{Mpc}\, (T/10~\mathrm{keV})^{1/2}$
(Evrard et al.\ 1996). We also required that the \ROSAT\/ exposure was
adequate for an accurate measurement of the surface brightness distribution
at large radii.  This requirement was implemented by the following objective
procedure. We fitted the power law index of the azimuthally averaged surface
brightness profile in the range $r>r_{180}/3$ and discarded all clusters
with a 1-$\sigma$ statistical uncertainty in their slope exceeding $\pm0.1$.
We also excluded clusters with double or very strongly irregular X-ray
morphology, because our analysis requires the assumption of reasonable
spherical symmetry. The excluded clusters were A754, Cyg-A, A1750, A2151,
A2197, A3223, A3556, A3558, A3560, A3562, A514, A548, S49-132, SC0625-536S,
A665, A119, A1763, A3266, and A3376. The 39 clusters satisfying all the above
criteria are listed in Table~\ref{tab:sample}.

The emission-weighted X-ray temperatures were compiled from the literature.
The main sources are \emph{ASCA} measurements by Markevitch et al.\ (1998)
and Fukazawa et al.\ (1998), both excluding the cooling flow regions, and
Mushotzky \& Scharf (1997), and a pre-\emph{ASCA} compilation by David et
al.\ (1993).  For three clusters without spectral data, we estimated
temperatures from the cooling flow corrected $L_x-T_x$ correlation derived
in Markevitch (1998); for two clusters, we adopted the $L_x-T_x$ temperature
estimates from Ebeling et al.\ (1996). The relative uncertainty of the
temperature estimates from the $L_x-T_x$ was assumed to be 25\% at the 68\%
confidence.

\section{\ROSAT\/ data reduction}
\label{sec:reduction}

\ROSAT\/ PSPC images were reduced using S.~Snowden's software (Snowden et
al.\ 1994). This software eliminates periods of high particle and scattered
solar backgrounds as well as 15-s intervals after turning the PSPC high
voltage on, when the detector may be unstable. Exposure maps in several
energy bands are then created using detector maps obtained during the
\ROSAT\/ All-Sky Survey that are appropriately rotated and convolved with
the distribution of coordinate shifts found in the observation. The exposure
maps include vignetting and all detector artifacts. The unvignetted particle
background is estimated and subtracted from the data to achieve a
high-quality flat-fielding even though the PSPC particle background is low
compared to the cosmic X-ray background. The scattered solar X-ray
background also should be subtracted separately, because, depending on the
viewing angle, it can introduce a constant background gradient across the
image. We eliminated most of Solar X-rays by simply excluding time intervals
when this emission was high, but the remaining contribution was also modeled
and subtracted.  The output of this procedure is a set of flat-fielded,
exposure corrected images in 6 energy bands, nominally corresponding to
0.2--0.4, 0.4--0.5, 0.5--0.7, 0.7--0.9, 0.9--1.3, and 1.3--2.0~keV (i.e.,
standard \emph{ROSAT} bands R2--R7). These images contain only cluster
emission, other X-ray sources, and the cosmic X-ray background. To optimize
the signal-to-noise ratio and to minimize the influence of Galactic
absorption, we used only the data above 0.7~keV\footnote{For five clusters
  (A2052, A2063, A2163, A3571, and MKW3S), we used the energy band
  0.9--2.0~keV to reduce the anomalously high soft background}. If the
cluster was observed in several pointings, each pointing was reduced
individually and the resulting images were merged.

To measure the cluster surface brightness distribution, we masked detectable
point sources and extended sources not related to the cluster. It is
ambiguous whether or not all sources should be excluded, because the angular
resolution varies strongly across the image and therefore a different
fraction of the background is resolved into sources. We chose to exclude all
detectable sources, and later checked that, with the exception of very
bright sources, the exclusion did not change our results.

The cluster surface brightness was measured in concentric rings of equal
logarithmic width; the ratio of the outer to inner radius of the ring was
equal to 1.1. We created both azimuthally averaged profiles and profiles in
six sectors with position angles $0^\circ-60^\circ$, ...,
$300^\circ-360^\circ$. The profile centroid was chosen at the cluster
surface brightness peak. The particular choice of the centroid can affect
the surface brightness profile in the inner region, especially for irregular
clusters. However, it does not change any results at large distances, which
was specifically checked. Therefore, we concluded that the simple choice of
the cluster centroid was sufficient for our regular clusters.

Finally, the cosmic X-ray background intensity was measured for each cluster
individually. Cluster flux often contributes significantly to the background
even at large distances from the center. We typically find that near
$r_{180}$, the cluster contributes around $5-20\%$ of the background
brightness.  Since $r_{180}$ can be quite close to the edge of the FOV, it
is often impossible to use any image region as a reference background
region.  Instead, we assumed that at large radii the cluster surface
brightness is a power law function of radius, and therefore the observed
brightness can be modeled as a power law plus constant background. Fitting
the data at $r>r_{180}/3$ with this model, we determined the background. We
checked that this technique provides the correct background value for
distant clusters where one can independently measure the background near the
PSPC edge. The fitted background value was subtracted from the data and its
statistical uncertainty included in the results presented below.

We have checked the flat-fielding quality using several \emph{ROSAT} PSPC
observations of ``empty'' fields. After exclusion of bright sources, as we
do in the analysis of clusters images, the difference in the background
level near the optical axis and near the FOV edge does not exceed $\sim
5\%$.  The $5\%$ background variations correspond to an additional
uncertainty of $\delta\beta\sim 0.03-0.04$ in $\beta$ (\S~\ref{sec:beta}) and
a 1--2\% uncertainty in the gas overdensity radius (\S~\ref{sec:r-t:obs}).

\begin{figure*}
  \centerline{\includegraphics[bb=43 589 571 739]{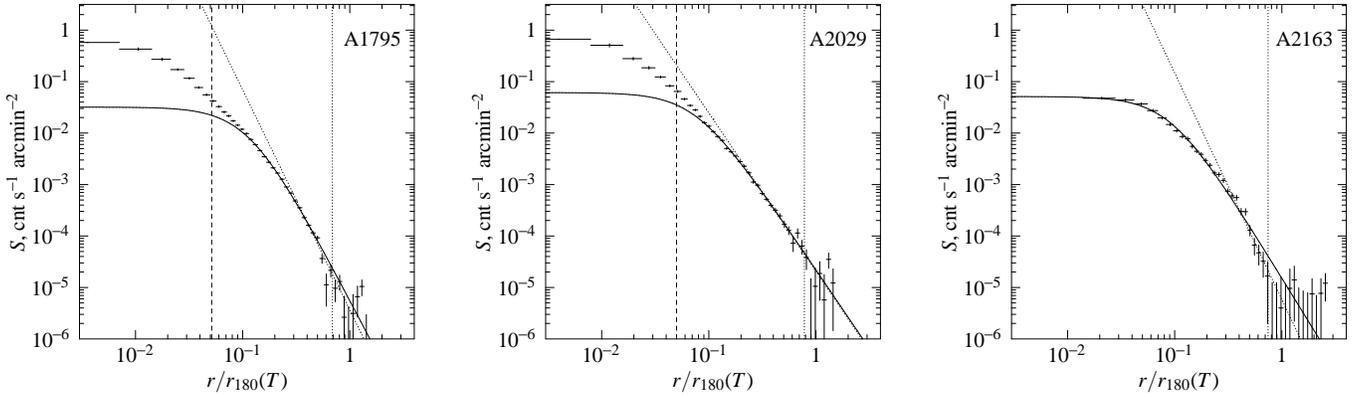}}
  \caption{Examples of surface brightness profiles. Vertical
    dashed lines show the cooling radius $R_{\rm cool}$. Vertical dotted
    lines show the radius of the mean baryon overdensity 1000, derived
    below. Solid lines show the $\beta$-model fit in the radial range $r>2
    R_{\rm cool}$. Dotted lines corresponds to the power-law fit for
    $0.3\r180<r<1.5\r180$.}
  \label{fig:prof:example}
\end{figure*}

\section{Surface Brightness Fits}
\label{sec:beta}

Cluster X-ray surface brightness profiles are usually modeled with the
$\beta$-model (Cavaliere \& Fusco-Femiano 1976) of the form
\begin{equation}
  \label{eq:beta:def}
  S(r) = S_0\left(1+r^2/r_c^2\right)^{3\beta-0.5},
\end{equation}
where $r$ is the angular projected off-center distance, and $S_0$,
$r_c$, and $\beta$ are free parameters. Jones \& Forman (1984, 1998)
fitted this equation to a large number of the {\em Einstein\/} IPC cluster
images. They find that the values of $\beta$ are distributed between $0.5$
and $0.8$ with the average ensemble value of $\langle\beta\rangle=0.6$. 
Jones \& Forman also find a mild trend of $\beta$ with the cluster
temperature in the sense that hotter clusters have larger $\beta$.

Cosmological cluster simulations typically predict steeper gas profiles,
$\beta \approx 0.8-1$ (e.g., Navarro et al.\ 1995), in contradiction with
the data. Bartelmann \& Steinmetz (1996) suggested that the observed values
of $\beta$ are underestimated because the surface brightness is saturated by
the background at large radii, where the brightness profile steepens.  The
accuracy of the $\beta$-model derived from the X-ray data is of great
importance because this model is widely used to derive the total gravitating
cluster mass via the hydrostatic equilibrium equation and to measure the gas
mass. Below we critically examine whether the $\beta$-model provides an
accurate description of the profiles in the wide range of radii, and also
whether the azimuthal averaging of the surface brightness of regular-looking
clusters can be justified. We also re-examine the previously reported
correlation of $\beta$ with temperature.

\subsection{Exclusion of Cooling Flows}

Many regular clusters have cooling flows which appear as strong peaks in the
surface brightness near the cluster center (Fabian 1994). The inclusion of
the cooling flow region in the $\beta$-model fit typically leads to small
values for the core-radius and $\beta$ and to a poor fit to the overall
brightness profile. Clearly, this region should be excluded from the fit if
an accurate modeling of the surface brightness at large radii is the goal.
Different strategies of the choice of the excluded regions can be found in
the literature. Jones \& Forman (1984) increased the radius of the excluded
region until the $\beta$-model fit provided an acceptable $\chi^2$. This
technique leads to different exclusion radii depending on the observation
exposure, cluster flux, and the radial bining of the surface brightness
profile.

A more physical approach would be to determine the radius beyond which gas
cooling cannot possibly be important, i.e.\ where the gas cooling time (see,
e.g.\ Fabian 1994) significantly exceeds the age of the Universe.  White,
Jones, \& Forman (1997) and Peres et al.\ (1998) provide the values of
$r_{\rm cool}$, the radius at which the cooling time equals
$1.3\times10^{10}\,$yr, for a large number of clusters, covering all but one
of the cooling flow clusters in our sample. We always excluded the region
$r<2\, r_{\rm cool}$, beyond which the cooling flow is unlikely to have any
effect on the surface brightness distribution.

\begin{figure*}
  \centerline{\includegraphics[bb=55 470 602 703]{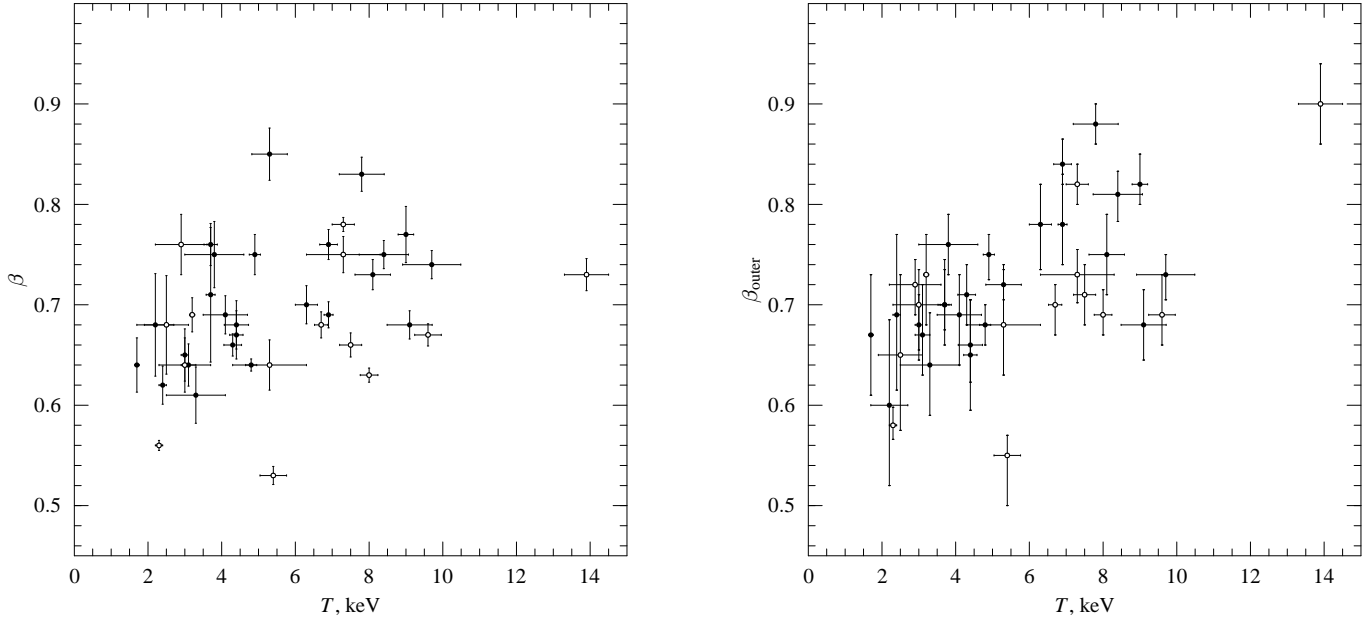}}
  \caption{Correlation of $\beta$ with temperature. The values of $\beta$
    are derived from the global fit and from the fit in the radial range
    $0.3\r180<r<1.5\r180$ in the left and right panels, respectively. Filled
    and open circles correspond to cooling flow and non cooling flow
    clusters, respectively.}
  \label{fig:beta-T}
\end{figure*}

\begin{table*}[htb]
  \footnotesize
  \begin{center}
    \caption{Surface Brightness Fitting and Gas Mass Results}
    \begin{tabular}{lcccccccc}
      \hline
      \hline
      \multicolumn{1}{c}{Name} & 
      \multicolumn{1}{c}{$\beta_{\rm outer}^{\rm (a)}$} &
      \multicolumn{1}{c}{$\beta_{rc, \rm out}^{\rm (b)}$} &
      \multicolumn{1}{c}{$\beta^{\rm (c)}$} &
      \multicolumn{1}{c}{$r_c^{\rm (c)}$} &
      \multicolumn{1}{c}{$R_{2000}^{\rm (d)}$} &
      \multicolumn{1}{c}{$R_{1000}^{\rm (d)}$} &
      \multicolumn{1}{c}{$\delta \beta_{\rm azi}^{\rm (e)}$} &
      \multicolumn{1}{c}{$\delta M_{\rm azi}^{\rm (f)}$} \\
\hline
   2A0335 & $0.68\pm0.03$ & $0.65$ & $0.65\pm0.03$ & ($0.08\pm0.08$) & $1.33\pm0.03$& $1.88\pm0.07$ & 0.04 & 0.09\\
     A133 & $0.76\pm0.03$ & $0.78$ & $0.75\pm0.03$ & ($0.37\pm0.05$) & $1.36\pm0.03$& $1.93\pm0.09$ & 0.09 & 0.16\\
    A1413 & $0.70\pm0.02$ & $0.67$ & $0.68\pm0.01$ & \phantom{(}$0.20\pm0.01$\phantom{)} & $1.66\pm0.10$& $2.43\pm0.15$ & 0.08 & 0.17\\
    A1651 & $0.78\pm0.04$ & $0.75$ & $0.70\pm0.02$ & ($0.26\pm0.03$) & $1.70\pm0.06$& $2.41\pm0.15$ & 0.23 & 0.18\\
    A1689 & $0.82\pm0.02$ & $0.79$ & $0.77\pm0.03$ & ($0.27\pm0.05$) & $1.76\pm0.07$& $2.46\pm0.15$ & 0.00 & 0.14\\
    A1795 & $0.88\pm0.02$ & $0.89$ & $0.83\pm0.02$ & ($0.39\pm0.02$) & $1.73\pm0.01$& $2.37\pm0.05$ & 0.12 & 0.09\\
    A2029 & $0.68\pm0.03$ & $0.67$ & $0.68\pm0.01$ & ($0.28\pm0.03$) & $2.02\pm0.05$& $2.90\pm0.12$ & 0.00 & 0.12\\
    A2052 & $0.67\pm0.04$ & $0.65$ & $0.64\pm0.02$ & ($0.10\pm0.05$) & $1.19\pm0.06$& $1.77\pm0.12$ & 0.04 & 0.12\\
    A2063 & $0.69\pm0.04$ & $0.68$ & $0.69\pm0.02$ & ($0.22\pm0.02$) & $1.19\pm0.05$& $1.71\pm0.14$ & 0.11 & 0.14\\
      A21 & $0.68\pm0.05$ & $0.69$ & $0.64\pm0.02$ & \phantom{(}$0.31\pm0.03$\phantom{)} & $1.42\pm0.08$& $2.08\pm0.17$ & 0.00 & 0.15\\
    A2142 & $0.73\pm0.02$ & $0.73$ & $0.74\pm0.01$ & ($0.42\pm0.03$) & $2.29\pm0.03$& $3.23\pm0.10$ & 0.05 & 0.11\\
    A2163 & $0.90\pm0.04$ & $0.89$ & $0.73\pm0.02$ & \phantom{(}$0.42\pm0.02$\phantom{)} & $2.48\pm0.09$& $3.42\pm0.17$ & 0.00 & 0.23\\
    A2199 & $0.68\pm0.02$ & $0.67$ & $0.64\pm0.01$ & ($0.14\pm0.01$) & $1.41\pm0.01$& $2.02\pm0.02$ & 0.07 & 0.07\\
    A2218 & $0.71\pm0.03$ & $0.70$ & $0.66\pm0.01$ & \phantom{(}$0.24\pm0.01$\phantom{)} & $1.59\pm0.05$& $2.33\pm0.08$ & 0.07 & 0.15\\
    A2255 & $0.73\pm0.03$ & $0.77$ & $0.75\pm0.02$ & \phantom{(}$0.55\pm0.02$\phantom{)} & $1.70\pm0.05$& $2.53\pm0.10$ & 0.18 & 0.06\\
    A2256 & $0.82\pm0.02$ & $0.82$ & $0.78\pm0.01$ & \phantom{(}$0.46\pm0.01$\phantom{)} & $1.88\pm0.01$& $2.69\pm0.03$ & 0.05 & 0.08\\
    A2382 & $0.72\pm0.03$ & $0.81$ & $0.76\pm0.03$ & \phantom{(}$0.47\pm0.03$\phantom{)} & $1.13\pm0.05$& $1.73\pm0.10$ & 0.04 & 0.22\\
    A2462 & $0.65\pm0.08$ & $0.67$ & $0.68\pm0.05$ & \phantom{(}$0.22\pm0.04$\phantom{)} & $0.86\pm0.11$& $1.30\pm0.23$ & 0.00 & 0.23\\
    A2597 & $0.66\pm0.04$ & $0.67$ & $0.68\pm0.02$ & ($0.18\pm0.04$) & $1.32\pm0.09$& $1.93\pm0.14$ & 0.13 & 0.15\\
    A2657 & $0.70\pm0.03$ & $0.75$ & $0.76\pm0.02$ & ($0.37\pm0.02$) & $1.22\pm0.03$& $1.76\pm0.07$ & 0.07 & 0.11\\
    A2717 & $0.60\pm0.08$ & $0.67$ & $0.68\pm0.05$ & ($0.07\pm0.08$) & $0.84\pm0.05$& $1.18\pm0.12$ & 0.00 & 0.13\\
    A3112 & $0.71\pm0.03$ & $0.69$ & $0.63\pm0.02$ & ($0.12\pm0.08$) & $1.53\pm0.05$& $2.17\pm0.13$ & 0.02 & 0.12\\
    A3301 & $0.70\pm0.04$ & $0.72$ & $0.64\pm0.03$ & \phantom{(}$0.29\pm0.03$\phantom{)} & $1.04\pm0.06$& $1.58\pm0.12$ & 0.09 & 0.15\\
    A3391 & $0.55\pm0.03$ & $0.54$ & $0.53\pm0.01$ & \phantom{(}$0.21\pm0.01$\phantom{)} & $1.41\pm0.06$& $2.25\pm0.18$ & 0.08 & 0.06\\
    A3571 & $0.78\pm0.04$ & $0.77$ & $0.69\pm0.01$ & ($0.27\pm0.02$) & $1.87\pm0.06$& $2.62\pm0.11$ & 0.04 & 0.07\\
     A400 & $0.58\pm0.02$ & $0.58$ & $0.56\pm0.01$ & \phantom{(}$0.18\pm0.01$\phantom{)} & $0.92\pm0.02$& $1.45\pm0.06$ & 0.01 & 0.11\\
     A401 & $0.69\pm0.02$ & $0.68$ & $0.63\pm0.01$ & \phantom{(}$0.27\pm0.01$\phantom{)} & $2.08\pm0.04$& $2.98\pm0.13$ & 0.06 & 0.17\\
    A4038 & $0.64\pm0.05$ & $0.63$ & $0.61\pm0.03$ & ($0.16\pm0.06$) & $1.15\pm0.07$& $1.64\pm0.11$ & 0.00 & 0.15\\
    A4059 & $0.65\pm0.05$ & $0.66$ & $0.67\pm0.02$ & ($0.22\pm0.05$) & $1.30\pm0.06$& $1.93\pm0.12$ & 0.05 & 0.14\\
     A478 & $0.81\pm0.02$ & $0.80$ & $0.75\pm0.01$ & ($0.31\pm0.03$) & $1.94\pm0.03$& $2.70\pm0.10$ & 0.07 & 0.09\\
     A496 & $0.75\pm0.02$ & $0.74$ & $0.70\pm0.02$ & ($0.25\pm0.02$) & $1.43\pm0.03$& $1.99\pm0.10$ & 0.00 & 0.08\\
     A539 & $0.73\pm0.04$ & $0.74$ & $0.69\pm0.02$ & \phantom{(}$0.25\pm0.01$\phantom{)} & $1.06\pm0.05$& $1.52\pm0.10$ & 0.32 & 0.23\\
     A644 & $0.75\pm0.04$ & $0.73$ & $0.73\pm0.02$ & ($0.24\pm0.02$) & $1.69\pm0.05$& $2.38\pm0.14$ & 0.00 & 0.13\\
      A85 & $0.84\pm0.02$ & $0.86$ & $0.76\pm0.02$ & ($0.40\pm0.02$) & $1.81\pm0.03$& $2.51\pm0.09$ & 0.11 & 0.18\\
     AWM4 & $0.69\pm0.08$ & $0.67$ & $0.62\pm0.02$ & ($0.11\pm0.01$) & $0.75\pm0.05$& $1.17\pm0.11$ & 0.00 & 0.14\\
     A780 & $0.71\pm0.03$ & $0.69$ & $0.66\pm0.01$ & ($0.12\pm0.03$) & $1.48\pm0.02$& $2.02\pm0.08$ & 0.19 & 0.13\\
    MKW3S & $0.70\pm0.04$ & $0.72$ & $0.71\pm0.07$ & ($0.30\pm0.10$) & $1.22\pm0.06$& $1.76\pm0.15$ & 0.00 & 0.11\\
     MKW4 & $0.67\pm0.06$ & $0.67$ & $0.64\pm0.03$ & ($0.18\pm0.02$) & $0.71\pm0.05$& $1.11\pm0.09$ & 0.15 & 0.12\\
  Tri Aus & $0.69\pm0.03$ & $0.69$ & $0.67\pm0.01$ & \phantom{(}$0.36\pm0.02$\phantom{)} & $2.29\pm0.04$& $3.22\pm0.11$ & 0.07 & 0.11\\
\hline
\end{tabular}

    \label{tab:results}
  \end{center}
$^{\rm(a)}$ --- results of the fit for the radial range
$0.3\r180<r<1.5\r180$ with core-radius fixed at $0.1\r180$.  $^{\rm(b)}$ ---
$\beta$ parameter in the $0.3\r180<r<1.5\r180$ range with core radius fixed
at a value derived from the fit for the entire radial range. Uncertainties
in $\beta_{rc, \rm out}$ and in $\beta_{\rm outer}$ are similar.
$^{\rm(c)}$ --- results over the entire radial range (excluding the cooling
flow); core-radius values for cooling flow clusters are given in
parentheses. $^{\rm(d)}$ --- radius of the mean gas overdensity
$\Delta=1000$ and 2000 relative to the background baryon density.
$^{\rm(e)}$ --- Azimuthal \emph{rms} variations of $\beta_{\rm outer}$ in
excess of the statistical variations. $^{\rm(f)}$ --- Azimuthal relative
\emph{rms} variations of the gas mass within $R_{1000}$, including
statistical noise.

\end{table*}

\subsection{Surface Brightness Slope}

For comparison with previous studies, the results of fitting the beta-model
to azimuthally averaged surface brightness profiles in the radius range
$2\,r_{\rm cool}$--$1.5\, r_{180}(T)$ for cooling flow, and $0$--$1.5\,
r_{180}(T)$ for non-cooling flow clusters\footnote{Cluster X-ray emission
  never has been detected to $1.5\, r_{180}(T)$.}, are presented in
Table~\ref{tab:results}. For cooling flow clusters, the best fit values of
core radius are often comparable to the radius of the excluded region;
therefore, the core radii cannot be reliably measured for those clusters.
The $\beta$-parameter, on the other hand, is measured very accurately, and
the $\beta$-model fits generally provide a very good description of the data
(see examples in Fig.~\ref{fig:prof:example}). The best-fit values of
$\beta$ are plotted versus the cluster temperature in Fig.~\ref{fig:beta-T}.
Similarly to Jones \& Forman (1998), we find that values of $\beta$ are
distributed over a narrow range $0.7\pm0.1$ for most clusters. However, the
distributions in our and Jones \& Forman samples are slightly offset. Jones
\& Forman find the average value $\langle\beta\rangle=0.6$, while all but
two our clusters have $\beta>0.6$. This difference is attributable in part
to different techniques of excising the cooling flows; but also, because of
the larger field of view and lower background, \ROSAT\/ data trace the
surface brightness to larger radius where the profiles often steepen (see
below).

Unlike, for example, clusters in the Jones \& Forman (1984, 1999) sample,
there are no hints of a correlation of $\beta$ with cluster temperature
(left panel in Fig.~\ref{fig:beta-T}). A careful examination shows that the
previously reported correlation of $\beta$ with temperature is due to small
values of $\beta\sim0.5$ for cool clusters with $T\sim 3$~keV, for which we
find significantly steeper profiles. Again, a likely explanation for this
discrepancy is the incomplete removal of cooling flows in the earlier
studies; a cooling flow, if not accounted for completely, biases $r_c$ and
$\beta$ low.


To determine the slope of the surface brightness profiles at large radii, we
fit the profiles in the same range of radii in virial coordinates, $0.3\,
r_{180}(T) < r < 1.5\, r_{180}(T)$. With this choice of radii, clusters of
different temperatures are compared in the same range of physical
coordinates. The core radius cannot be determined from this fit, and so we
fixed its value at either $0.1\,r_{180}$ or the value derived from the fit
for the entire radial range. Because the value for core-radius is typically
much smaller than the inner radius of the data, these modelings are equivalent
to fitting the power law relation, $S\propto r^{-6\beta+1}$. The values of
$\beta_{\rm outer}$ are listed in Table~\ref{tab:results} and plotted versus
cluster temperature in Fig.~\ref{fig:beta-T}.

The slopes in the outer parts in many clusters are slightly steeper than
those given by the $\beta$-model.  The extreme case is A2163, where $\beta$
changes by 0.17. The surface brightness profile of this cluster shows a
clear steepening at $r>0.3\,r_{180}(T)$ (Fig.~\ref{fig:beta-T}). Although
this cluster is probably a merger (see discussion in Markevitch et al.\ 
1996), the same steepening in the surface brightness is seen in all but one
of the $60^\circ$ sectors. However, the typical change of $\beta$ in the
outer parts is much smaller, $\Delta\beta\approx0.05$, and only marginally
significant in most clusters. Thus, a strong steepening of the gas density
distribution at large radius suggested by Bartelmann \& Steinmetz (1996) is
excluded.

There is some indication of a positive correlation of $\beta_{\rm outer}$
with temperature. This is mainly due to a group of 5 hot, $T=6-10$~keV,
clusters with $\beta_{\rm outer}>0.8$, and a strong steepening of the
surface brightness profile in the hottest cluster A2163. However, as is seen
from Fig.~\ref{fig:beta-T}, the possible change of slope is well within the
scatter at high temperatures. In any case, the change of slope is small,
from $\beta\approx0.67$ for 3~keV clusters to $\beta\approx 0.8-0.85$ for
10~keV clusters.

\subsection{Azimuthal Variations of the Surface Brightness}
\label{sec:beta:azi}

Cluster X-ray surface brightness is often described by a radial profile (as
in the previous sections). It is important to determine how accurate this
description is in the outer region.  We divide the clusters into sectors
0$^\circ$--60$^\circ$, ..., 300$^\circ$--360$^\circ$, and determine
$\beta_{\rm outer}$ in the radial range $0.3\, r_{180}(T) < r < 1.5\,
r_{180}(T)$ in each sector separately. Azimuthal variations of $\beta_{\rm
  outer}$ would indicate an asymmetric cluster.

\centerline{\includegraphics[bb=55 469 326 732]{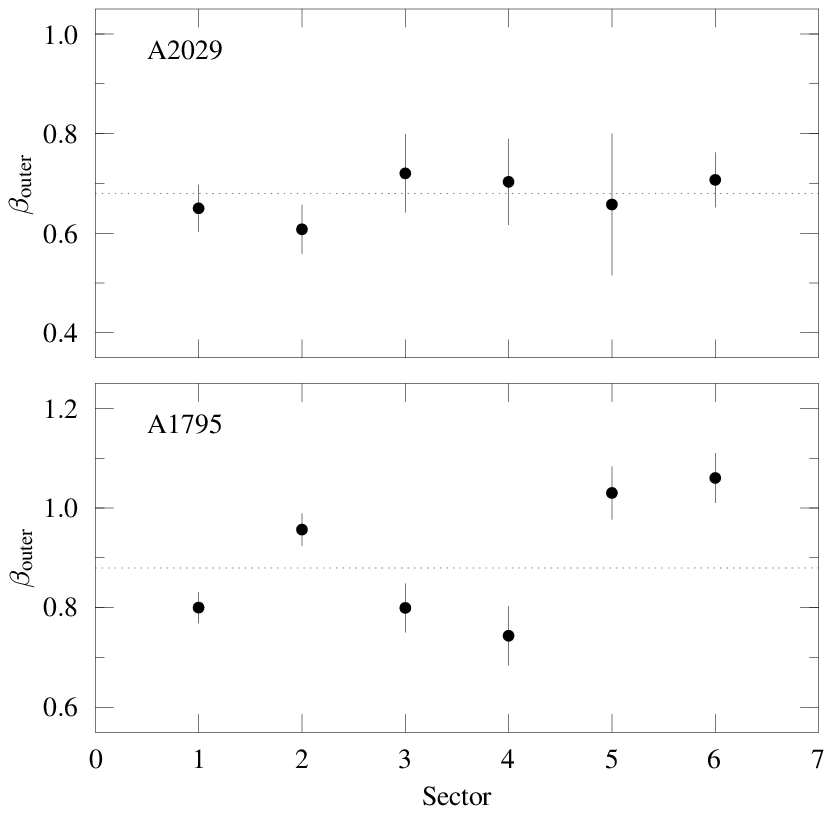}}
\figcaption{Azimuthal variations of $\beta$ in the outer part of A2029 and
  A1795.\label{fig:beta:azi}}
\smallskip

The sample appears to have clusters from very regular ones, such as A2029,
to those which display statistically significant azimuthal variations of the
slope, such as A1795 (Fig.~\ref{fig:beta:azi}). However, the amplitude of
variations is typically not very large. The azimuthal {\em rms\/} variations
of $\beta_{\rm outer}$ in excess of the statistical noise level are listed
for all clusters in Table~\ref{tab:results}.  In most cases, these
variations are below 0.1, and in many cases are dominated by a strong
deviation in just one sector. We conclude that the azimuthal averaging of
the surface brightness in the cluster outer parts can be justified. We will
return to the issue of azimuthal averaging in the discussion of gas mass
distribution below.

\section{Gas Mass Distribution}

The X-ray surface brightness distribution in clusters receives much
attention because it can be rather precisely converted to the distribution
of hot gas.  Determination of the gas mass distribution is also a goal of
our study. Below we briefly review techniques used to derive the gas density
and present the results for our sample.

\begin{figure*}
  \centerline{\includegraphics[bb=55 470 602 703]{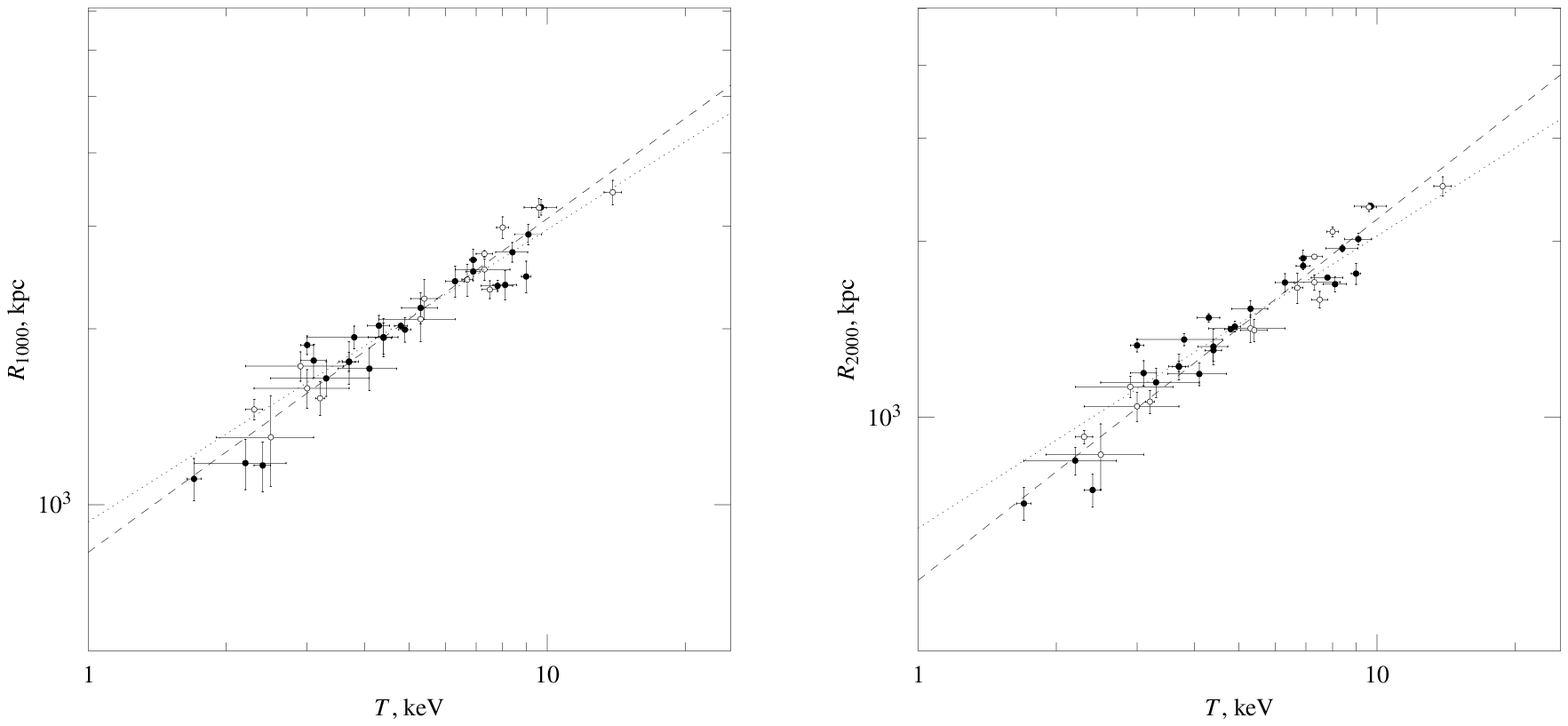}}
  \caption{Correlation of $R_{1000}$ and $R_{2000}$ with
    temperature. Dashed lines show the best power law fit, $R_{1000}\propto
    T^{0.57}$ and $R_{2000}\propto T^{0.61}$, while the dotted lines show
    $R\propto T^{0.5}$ fits. Filled and open circles correspond to cooling
    flow and non cooling flow clusters, respectively.}
  \label{fig:rover}
\end{figure*}

\subsection{Conversion of Surface Brightness to Gas Mass}

Under the assumption of spherical symmetry, the observed surface brightness
profile can be converted to the emissivity profile. The latter is then
easily converted to the gas density profile, because the X-ray emissivity of
hot, homogeneous plasma is proportional to the square of the density and, in
the soft X-ray band, depends only very weakly on temperature (e.g.,
Fabricant, Lecar, \& Gorenstein 1980, and \S\ref{sec:worries:t} below).

There are two main techniques to convert the observed surface brightness
profile to the emissivity profile under the assumption of spherical
symmetry. The first method is to fit an analytical function to the surface
brightness profile $S(r)$ and then deproject the fit using the inverse Abell
integral (e.g., Sarazin 1986). For the $\beta$-model surface brightness fit
(eq.~\ref{eq:beta:def}), the conversion is particularly simple (Cavaliere \&
Fusco-Femiano 1976, Sarazin 1986).

The second widely used technique is the direct deprojection of the data
without using an analytical model (Fabian et al.\ 1981, Kriss, Cioffi, \&
Canizares 1983). Briefly, one assumes that the emissivity is uniform within
spherical shells corresponding to the surface brightness profile annuli. The
contribution of the outer shells to the flux in each annuli can be
subtracted, and the emissivity in the shell calculated, using simple
geometrical considerations. This method has an important advantage over the
$\beta$-fit, in that no functional form of the gas distribution is assumed
and realistic statistical uncertainties at each radius are obtained.
Although we generally find very little difference between the deprojection
and $\beta$-fit methods, we use the deprojection technique as the preferred
one.

Once the distribution of emissivity (in units of flux per volume) is known,
it can be converted to the distribution of gas mass as follows. The
emissivity is multiplied by the volume of the spherical shell to obtain the
total flux from this shell. Assuming that the gas temperature is constant at
all radii, we use the Raymond \& Smith (1977) spectral code to find the
conversion coefficient between the flux and the emission measure integral,
$E=\int n_e n_p \,dV$, given the plasma temperature, heavy metal abundance,
cluster redshift, and Galactic absorption. Metal abundance has virtually no
effect on the derived gas mass at high temperatures, which is the case for
our clusters; we assume that it is 0.3 of the Solar value for all clusters.
For this metal abundance, $n_e/n_p=1.17$, and the gas density is
$\rho_g=1.35 \,m_p n_p$. The gas mass in the shell is $m_g = m_p
(1.56\,E\,V)^{1/2}$, where $V$ is the volume of the shell.  Given the
observed flux, the derived gas mass scales with distance to the cluster as
$d^{5/2}$.

\subsection{Correlation of the Baryon Overdensity Radius with Temperature}
\label{sec:r-t:obs}

As was pointed out in \S1, simple theory predicts a tight correlation
between the radius at a fixed baryon overdensity relative to the
background density of baryons and the temperature in the form $R\propto
T^{0.5}$. Since most baryons in clusters are in the form of hot gas, and the
gas mass is easily measured from the X-ray data, this correlation can be
tested observationally.

We use the deprojection technique to determine the enclosed gas mass as a
function of radius. The baryon overdensity is calculated as the ratio of
the enclosed mass and $(4\pi/3)\rho_0 R^3 (1+z)^3$, where $\rho_0$ is the
present day background density of baryons derived from primordial
nucleosynthesis, and $z$ is the cluster redshift. We adopt the value
$\rho_0=2.85\times10^9$~$M_\odot$~Mpc$^{-3}$ (Walker et al.\ 1991); a
different value of the background baryon density (e.g., a recent
determination by Burles \& Tytler 1998) would have no effect on our results
except for scaling the reported overdensities.

Previous studies of the baryonic contents in clusters indicated that baryons
contribute $\sim15-20\%$ of the total cluster mass (for $h=0.5$); if this
ratio is representative of the Universe as a whole, it corresponds to a
cosmological density parameter $\Omega_0=0.2-0.3$ (White et al.\ 1993, David
et al.\ 1995, Evrard 1997). With this range for $\Omega_0$, the two commonly
referenced values of the dark matter overdensity $\delta=180$ and 500
relative to the critical density correspond to gas overdensities
$\Delta_g=600-1000$ and 1500--2500, respectively. Therefore, we determine
the radii at which the mean enclosed hot gas density is 1000 and 2000 above
the baryon background; these radii are denoted $R_{1000}$ and $R_{2000}$
hereafter.

For a wide range of gas temperatures, from 1.5 to 10 keV, the gas mass
corresponding to the fixed \emph{ROSAT} flux changes by only 4\% if metal
abundance is $a=0.3$ Solar, and by 10\%, if $a=0.5$. The corresponding
variations of the gas overdensity radius are approximately $2\%$ and 5\%.
Therefore, the values of $R_{1000}$ and $R_{2000}$, as derived from the
\emph{ROSAT} data, are practically independent of the gas temperature.

The measured radii $R_{1000}$ and $R_{2000}$ are plotted versus cluster
temperature in Fig.~\ref{fig:rover}. Note that $R$ and $T$ are measured
essentially independently, as opposed, for example, to the baryon fraction
or total mass that involves mass estimates that use the gas temperature.
The correlation is very tight and close to the theoretically expected
$R\propto T^{0.5}$. Note that even A3391, the cluster with an anomalously
flat surface brightness profile, is quite close to the observed correlation.
We fit power laws to the $R-T$ relation using the bisector modification
of the linear regression algorithm that allows for intrinsic scatter and
non-uniform measurement errors, and treats both variables symmetrically
(Akritas \& Bershady 1996 and references therein). The confidence intervals
were determined using bootstrap resampling (e.g., Press et al.\ 1992). The
best fit relations are
\begin{eqnarray}
  \lg R_{1000} &=& (0.569\pm0.043)\, \lg T + 2.918 \nonumber\\
  \lg R_{2000} &=& (0.615\pm0.042)\, \lg T + 2.720,
\end{eqnarray}
where radii are in kpc and temperatures are in keV, and uncertainties are
$68\%$ confidence. For any given temperature, the average scatter in
$R_{1000}$ is only 6.5\%, and $\sim 7\%$ for $R_{2000}$. This is comparable
to the scatter of the dark matter overdensity radius $r_{500}$ in simulated
clusters (Evrard et al.\ 1996).  Even though the best fit slopes formally
deviate from the expected value of 0.5 by $2-3\sigma$, the difference
between the best fit and the $R\propto T^{0.5}$ relation is within the
scatter in the data (Fig.~\ref{fig:rover}).

\begin{figure*}[htb]
  \centerline{\includegraphics[bb=49 565 569 716]{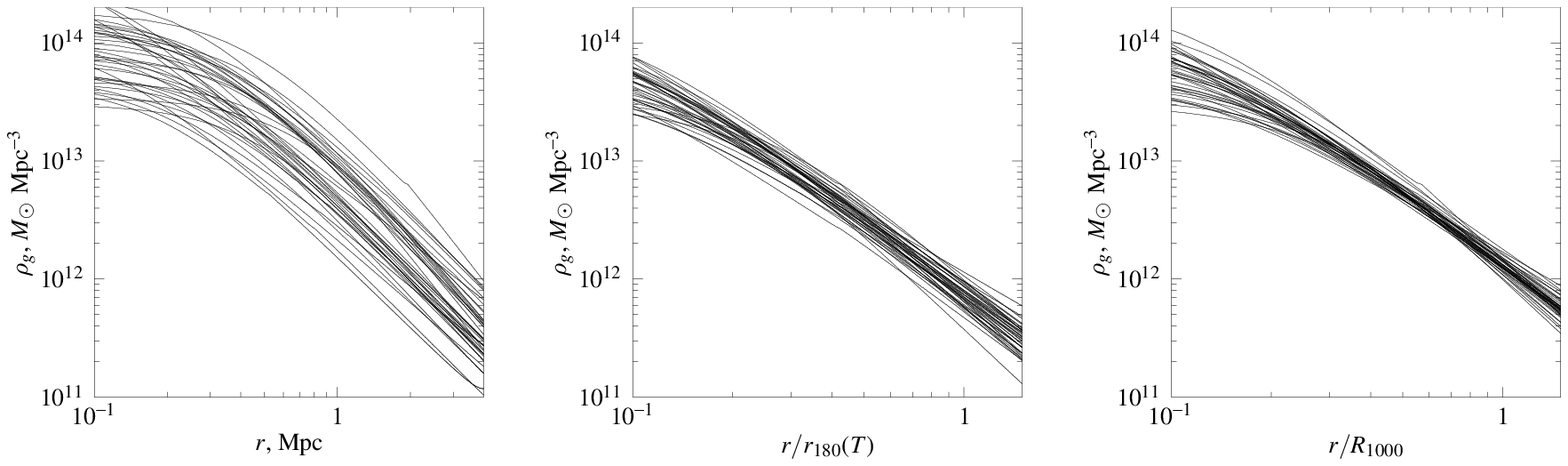}}
  \caption{Raw and scaled gas density profiles. For readability,
    density profiles are calculated from the $\beta$-model fits instead of
    our preferred deprojection technique.}
  \label{fig:comp:gas:prof}
\end{figure*}

A tight correlation of the baryon overdensity radius with temperature
suggests that the gas density profiles in the outer parts of clusters are
similar, when appropriately scaled. Figure~\ref{fig:comp:gas:prof} shows the
gas density profiles plotted as a function of radius in Mpc, in units of
$r_{180}(T)$, and in units of $R_{1000}$. No density scaling was applied. As
expected, density profiles display a large scatter if no radius scaling is
applied, since we are comparing systems of widely different masses. The
density scatter at large radius becomes small (the entire range is $\approx
\pm40\%$) when radii are scaled to $r_{180}(T)$.  This scatter is close to
that of the dark matter density in simulated clusters. The scatter is
particularly small when profiles are scaled to the overdensity radius
$R_{1000}$. One can argue that in this case, the scatter is artificially
suppressed because the scaling depends on the density. However, the scatter
remains small over a rather wide radial range; also, the same critique
applies when dark matter profiles of simulated clusters are plotted in
virial coordinates.

To conclude, gas density profiles show a high degree of similarity, both in
terms of enclosed mass and shape, when the radius is scaled to either the
virial radius estimated from the gas temperature or the fixed gas overdensity
radius.  In the next section, we discuss some uncertainties which affect the
measurement of the gas mass distribution.

\section{Discussion of Gas Mass Uncertainties}
\label{sec:worries}

\subsection{Sample Selection and Spherical Symmetry}

To calculate gas mass from the observed X-ray surface brightness, we, and
most other studies, assume that the cluster is spherically symmetric. It is
desirable to check that this assumption is adequate.  If the substructure
had a strong effect on the derived gas mass, the mass calculated using
surface brightness profiles in different cluster sectors would be
substantially different when the substructure is seen in projection. Because
of random orientations, substructure in projection occurs more often than
along the line of sight. Therefore, the azimuthal variations of the gas mass
can be used to place limits on the 3-dimensional deviations from the
spherical symmetry. To look for this effect, we calculated gas masses using
surface brightness profiles in six sectors in all clusters.  The {\em rms\/}
azimuthal gas mass variations within $R_{1000}$ are listed in
Table~\ref{tab:results}.  For most clusters, these variations are on the
level of $\sim 10-15\%$, including statistical scatter. Since we find that
mass variations in projection are small, this indicates that they also are
small in three-dimensional space.

It can be argued that only small azimuthal mass variations are found because
clusters with substructure were excluded from the sample. Moreover, such
selection might lead to a preferential selection of clusters having
substructure along the line of sight which is invisible in the images. As a
result, our gas distribution measurements might be seriously biased, because
we do not average over different cluster orientations.

These arguments are countered by the following considerations. We excluded
only three cooling flow clusters, \mbox{Cyg-A}, A3558, and A1763, on the
basis of strong substructure, compared to 25 such clusters in the sample
(Table~\ref{tab:sample}). Therefore, our cooling flow subsample, which
comprises two thirds of the total sample, should be unbiased with respect to
substructure selection. Since there is no obvious difference between cooling
and non-cooling flow clusters either in terms of gas mass
(Fig.~\ref{fig:rover}) or surface brightness fits (Fig.~\ref{fig:beta-T}),
our subsample of non-cooling flow clusters also is unlikely to have
significant substructure along the line of sight.

\subsection{Temperature Structure}
\label{sec:worries:t}

{\em ASCA}\/ measurements suggest that, at least in hot clusters, the gas
temperature gradually declines with radius reaching $\sim 0.5$ of the
average value at $r=0.5\,r_{180}$ (Markevitch et al.\ 1998). Because of
strong line emission, calculation of the gas density from \ROSAT\/ flux is
uncertain for cool clusters, if precise temperatures and metal abundances
are unknown.  If such cool ($T\sim2$~keV) clusters have declining
temperature profiles, our gas masses will be affected, because we assume
isothermality.  Fortunately, the effect is not very strong. We tested this
by simulating the $T=0.5$~keV Raymond-Smith plasma with heavy metal
abundance, $a$, in the range 0.1--0.5 of the Solar value, and converting the
predicted \ROSAT\/ PSPC flux in the 0.5--2~keV band back to gas mass using
the $T=2$~keV, $a=0.3$ spectral model. The mass was underestimated by 20\%
for $a=0.1$ and overestimated by $35\%$ for $a=0.5$.  For the input spectrum
with $T=1$~keV, the mass error was in the range $\pm15\%$. The effect of
temperature decline on the {\it enclosed}\/ gas mass is smaller, because a
significant mass fraction is contained within the inner, hotter regions. The
error in the gas overdensity radius determination is still smaller because
overdensity is a very strong function of radius (for example, $\Delta_g
\propto r^{-2}$ for $\beta=2/3$).

\subsection{Cooling Flows}
\label{sec:worries:cf}

The presence of a cooling flow leads to an overestimate of the gas density
near the cluster center, if one assumes that the gas is single-phase and
isothermal. However, the enclosed gas mass at large radius is little
affected, because most of gas mass lies at large radii. For example, in
A1795, the cluster with one of the strongest cooling flows, only 2.7\% of
the gas mass inside $R_{1000}$ is within the cooling radius and 9\% of the
mass is within $2\,r_{\rm cool}$, if one assumes that the cooling flow is
single-phase and isothermal. Even if the mass within $2\,r_{\rm cool}$ is
overestimated by 100\% because of these incorrect assumptions, the total gas
mass is overestimated by only 10\%, and $R_{1000}$ is overestimated by only
3\%.  The true errors are likely to be smaller.

The presence of a cooling flow also leads to an underestimate of the
emission-weighted temperature (underestimation here is relative to the
absence of radiative cooling, the assumption usually made in theory and
simulations).  For example, Markevitch et al.\ (1998) find that in several
clusters, the temperature increases by up to 30\% when the cooling flow is
excised. This temperature error produces almost no to errors in gas masses,
but can introduce an additional scatter in the $R-T$ correlation, or in the
gas density profiles scaled to $r_{180}(T)$. We used temperatures from
Markevitch et al.\ for which cooling flows were excised, for all clusters
with strong cooling flows, except 2A0335 and A1689. Allen and Fabian (1998)
find that the temperature increase in A1689 when the cooling flow is modeled
as an additional spectral component is small, $\sim 5\%$. Cooling flows in
other clusters in our sample are not very strong, and simple
emission-weighted temperatures should be sufficiently accurate.

\section{Discussion}

\subsection{Applicability of the $\beta$-model}

We have found above that the slope of the surface brightness profile in the
outer part of clusters [$0.3\,r_{180}-r_{180}$] is slightly steeper than the
slope of the $\beta$-model fit in the entire radial range (excluding the
cooling flow region). Thus, the $\beta$-model does not describe the gas
distribution precisely.  However, deviations from the $\beta$-model are
small and do not lead to significant errors in the total mass or gas mass.
Consider the extreme case of A2163, where the global $\beta$-value is 0.73
but beyond a radius of $0.3\,r_{180}(T)$, the profile slope steepens to
$\beta=0.9$.  Such a change of $\beta$ leads to a 24\% increase of the total
mass calculated from the hydrostatic equilibrium equation; this is smaller
than other uncertainties (Markevitch et al.\ 1996). The gas masses within
$R_{1000}$ calculated from the global $\beta$-model and from the exact
surface brightness profile differ by 20\%. In most clusters, where $\beta$
typically changes by $\approx 0.05$, the effect on the total and gas mass is
much smaller.

\subsection{$R-T$: The First ``Proper'' Scaling for Baryons}

The scaling relations involving hot gas in clusters established previously
show significant deviations from the theoretically expected relations. The
most notable example is the luminosity-temperature correlation. From the
virial theorem relation $M_{\rm tot}\propto T^{3/2}$, and the assumptions of
constant baryon fraction and self-similarity of clusters, one expects
$L_{x}\propto T^2$ while the observed relation is closer to $L_x \propto
T^3$ (David et al.\ 1993). The current consensus is that additional physics,
such as preheating of the intergalactic medium or the feedback from galaxy
winds/supernovae or shock heating of the IGM has important effects on the
X-ray luminosities (see Cavaliere, Menci \& Tozzi 1997 and references
therein). These processes are still uncertain, and for example, prevent the
use of the evolution of the cluster luminosity function as a cosmological
probe.

Another example of the deviations of cluster baryon scaling from theoretical
expectations is the relation between the cluster size at a fixed X-ray
surface brightness and temperature (Mohr \& Evrard 1997). Mohr \& Evrard
find $R_I \propto T^{0.9\pm0.1}$ from the observations, while their
simulated clusters show $R_I \propto T^{0.7}$. After inclusion of feedback
from galaxy winds to the simulations, Mohr \& Evrard were able to reproduce
the observed size-temperature relation. Note that the surface brightness
threshold used by Mohr \& Evrard was selected at a high level, so that the
derived size $R_I$ was only $\approx 0.3$ of the cluster virial radius.

The scaling between the radius of a fixed gas overdensity and temperature,
$R \approx \mathrm{const} \times T^{1/2}$, presented here is, to our
knowledge, the first observed scaling involving only cluster baryons that is
easily understandable theoretically (\S1). The crucial difference between
the luminosity-temperature and Mohr \& Evrdard's size-temperature relation
and our scaling is that we use cluster properties at large radius, where
most of the mass is located, while the $L-T$ and $R_I-T$ relations are based
on properties of the inner cluster regions. Our findings thus suggest that
any processes required to explain the observed $L-T$ and $R_I-T$ relations
affect only central cluster parts and are not important for the gas
distribution at large radii. 

\subsection{Limit on the Variations in the Baryon Fraction}

Simulations predict that the total mass within a radius of fixed overdensity
scales as $M_{\rm tot}\propto T^{3/2}$ (Evrard et al.\ 1996).  Our observed
scaling between the gas overdensity radius and $T$ is consistent with
$R\propto T^{1/2}$, or equivalently $M_{\rm gas}\propto T^{3/2}$.
Therefore, if the simulations are correct, $M_{\rm gas}/M_{\rm tot}$ does
not depend on the cluster temperature. Since hot gas is the dominant
component of baryons in clusters, the baryon fraction within a radius of
fixed overdensity is constant for all clusters. To be more precise, the
best-fit relation $R_{1000}\propto T^{0.57}$ corresponds to a slowly varying
gas fraction $M_{\rm gas}/M_{\rm tot} = T^{0.2}$. However, the stellar
contribution can reduce this trend, because stars contribute a greater
fraction of the baryon mass in low-temperature clusters (David et al.\ 
1990, David 1997).

The small observed scatter around the mean $R-T$ relation can be used to
place limits on the variations of the baryon fraction between clusters of
similar temperature. At large radius, the mean gas overdensity is $\Delta_g
\propto r^{-3\beta}$. Therefore, the $\sim 7\%$ observed scatter in radius
at the given $\Delta$ corresponds to a $3\beta\times7\%$ scatter in
overdensity at the given radius. Assuming that the total cluster mass is
uniquely characterized by the temperature, the scatter in $M_{\rm
  gas}/M_{\rm tot}$ is 14--18\%, \emph{including} the measurement
uncertainties.  The small scatter indicates that the baryon fraction in
clusters is indeed universal. There is also an intrinsic scatter in the
$M_{\rm tot}-T$ relation, which is 8\%--15\% in simulated clusters (Evrard
et al.\ 1996); if the deviations of the total mass and gas mass from the
average value expected for the given temperature are not anti-correlated,
the scatter of the baryon fraction is reduced still further.  Thus, our
results provide further observational support for measurements of $\Omega_0$
from the baryon fraction in clusters and the global density of baryons
derived from primordial nucleosynthesis.

\subsection{Similarity of Gas Density Profiles}

The gas density profiles plotted in virial coordinates, i.e., radius scaled
by either $r_{180}(T)$ or $R_{1000}$, are very similar, both in slope and
normalization (Fig.~\ref{fig:comp:gas:prof}). The similarity of the gas
density slopes in the outer parts of clusters also is evident from the
relatively small scatter of $\beta$-values in Fig.~\ref{fig:beta-T}. Most
clusters have $0.65<\beta_{\rm outer}<0.85$, which corresponds to gas
density falling with radius between $r^{-1.95}$ and $r^{-2.55}$.

The average gas density, $\rho_{\rm g} \sim r^{-2.25}$, is significantly
more shallow than the universal density profile of the dark matter halos
found in numerical simulations, $\rho_{\rm dm}\sim r^{-2.7}$ between
$0.3\,r_{180}$ and $r_{180}$ (Navarro et al.\ 1995). Moreover, if gas in
this radial range is in hydrostatic equilibrium and isothermal, a power law
function of gas density with radius implies $\rho_{\rm dm}\sim r^{-2}$.
Under the hydrostatic equilibrium assumption, the average gas polytropic
index $\gamma\sim1.3$, or equivalently $T\propto r^{-0.7}$, is required for
the total mass to follow the Navarro et al.\ distribution.  Interestingly,
this is quite close to the temperature profile observed in many clusters
within $0.5\,r_{180}$ (Markevitch et al.\ 1998).

\subsection{Comparison with Other Works}

After this paper was submitted, we learned about works of Mohr, Mathiesen \&
Evrard (1999) and Ettori \& Fabian (1999) who also studied the hot gas
distribution in large cluster samples. We briefly discuss some aspects of
these works that are in common with our study.

Both Mohr et al.\ and Ettori \& Fabian derive cluster $\beta$'s from a
global fit. Ettori \& Fabian exclude central 200~kpc in the cooling flow
clusters; they find the global $\beta$'s in the range 0.6--0.8, in agreement
with our results. Mohr et al.\ fit the cooling flow region with an
additional $\beta$-model component, but force the same $\beta$ for the
cluster and cooling flow components.  Their values of $\beta$ for
cooling flow clusters are often flatter than ours (e.g., they derive
$\beta=0.66\pm0.03$ for A85, while our value is 0.76); most likely, this is
due to the difference in fitting procedures.

Mohr et al.\ find a tight correlation of the cluster temperature with the
hot gas mass within $r_{500}$ (estimated as $r_{500} = C\times T^{1/2}$) in
the form $M_{\rm gas}\propto T^{1.98\pm0.18}$.  Because the gas mass and gas
overdensity radius are related as $M_{\rm gas} = {\rm const} \times R^3$,
the $M_{\rm gas}-T$ and our $R_{1000}-T$ correlations are almost equivalent.
However, our correlation corresponds to a flatter $M_{\rm gas}-T$ relation,
$M \propto T^{1.71\pm0.13}$, closer to the theoretically expected slope of
1.5. There is an important difference of our and Mohr et al.\ approaches.
While in our method, $R_{1000}$ and $T$ are measured essentially
independently, the Mohr et al.\ measurement of the gas mass does depend on
$T$ through $r_{500}$. Since $r_{500} \propto T^{1/2}$ and typically $M_{\rm
  gas}(<r) \sim r$, their method would find $M_{\rm gas} \propto T^{1/2}$
even if gas profiles of all clusters are the same. This effect may introduce
a bias which is responsible for a slightly steeper $M_{\rm gas}-T$ relation
in Mohr et al.

Mohr et al.\ and Ettori \& Fabian (for low-redshift clusters) find that the
values of gas fraction in hot clusters are distributed in a relatively
narrow range, $f_{\rm gas}\sim 0.2\pm0.04$. Our tight correlation of
$R_{1000}$ and $T$ also implies a low, $\sim 15\%$ scatter in $f_{\rm gas}$
(see above).

\section{Summary}

We have carried out a detailed analysis of the surface brightness
distributions of a sample of 25 cooling flow clusters and 14
non-cooling flow clusters. Since the bulk of the cluster gas mass, and
hence the luminous cluster baryons, reside at large radii, we have
focussed on the properties of  the gas profile at large
radii

The cluster profiles, from $0.3 r_{180}$ to $r_{180}$ can be accurately
characterized as a single power law with $\beta=0.65-0.85$.  These outer
profiles are steeper by about 0.05 in $\beta$ on average than profiles fit
using the entire surface brightness profile (but excluding the cooling flow
region). This indicates that the $\beta$-model does not describe the surface
brightness profiles precisely.

The previously reported correlation of increasing $\beta$ with increasing
temperature (steepening profiles with increasing temperatures) is only
weakly present in our data. This difference arises primarily because the low
\emph{ROSAT} background allows us to detect clusters to near the virial
radius where they exhibit more similar profiles than in the central part,
often dominated by the cooling flow.


We find a very precise correlation of the radius, corresponding to a fixed
baryon overdensity, with gas temperature which is consistent with that
theoretically predicted from the virial theorem. For example, the radius at
which the mean baryon overdensity is 1000 is best fit as a function of
temperature as $R_{1000} \propto T^{0.57\pm0.04}$ and is consistent within
the scatter with the theoretically expected relation $R \propto T^{0.5}$.

The observed scatter in the correlation of $R_{1000}$ vs.\ $T$ is small.
Quantitatively, for any given temperature the average scatter in $R_{1000}$
is approximately 7\%. This corresponds to a scatter in $M_{\rm gas}/M_{\rm
  tot}$ at the same radius of less than 20\%, which includes any intrinsic
variation as well as measurement errors.

At large radii, cluster gas density distributions are remarkably similar
when scaled to the cluster virial radius ($r_{180}$) and they are
significantly shallower than the universal profile of dark matter density
found in simulations (Navarro et al.\ 1995). However, for gas in hydrostatic
equilibrium, the temperature profile found by Markevitch et al.\ (1998)
combined with the gas density profiles observed for our sample imply a dark
matter distribution quite similar to the universal one found in numerical
simulations.

\acknowledgements

M.\ Markevitch is thanked for careful reading of the manuscript. This work
was supported by the CfA postdoctoral fellowship.

\end{document}